\documentclass[journal]{IEEEtran}

\pdfoutput=1

\usepackage[T1]{fontenc}
\usepackage[utf8]{inputenc}
\usepackage{textcomp} 

\usepackage{amsmath,amsfonts,amssymb}
\usepackage{algorithmic}
\usepackage{array}
\usepackage[caption=false,font=normalsize,labelfont=sf,textfont=sf]{subfig}
\usepackage{textcomp}
\usepackage{stfloats}
\usepackage{url}
\usepackage{verbatim}
\usepackage{graphicx}

\usepackage[hidelinks]{hyperref}
\usepackage[numbers]{natbib}
\def\BibTeX{{\rm B\kern-.05em{\sc i\kern-.025em b}\kern-.08em
    T\kern-.1667em\lower.7ex\hbox{E}\kern-.125emX}}
\usepackage{balance}

\usepackage{booktabs}
\usepackage{longtable}
\usepackage{calc} 

\usepackage{url}
\graphicspath{{imgs/}}
\makeatletter
\@ifpackageloaded{subfig}{}{\usepackage{subfig}}
\@ifpackageloaded{caption}{}{\usepackage{caption}}
\AtBeginDocument{%

}
\AtBeginDocument{%

}
\newcounter{pandoccrossref@subfigures@footnote@counter}
{\end{figure}%
\addtocounter{footnote}{-\value{pandoccrossref@subfigures@footnote@counter}}
\@for\f:=\global@pandoccrossref@subfigures@footnotes\do{\stepcounter{footnote}\footnotetext{\f}}%
\gdef\global@pandoccrossref@subfigures@footnotes{}}
\@ifpackageloaded{float}{}{\usepackage{float}}
\floatstyle{ruled}
\@ifundefined{c@chapter}{\newfloat{codelisting}{h}{lop}}{\newfloat{codelisting}{h}{lop}[chapter]}
\floatname{codelisting}{Listing}

\makeatother

\def\[#1\]{%
  \begin{equation}#1\end{equation}%
}

\date{\today}

\begin{document}

\title{Towards detecting the pathological subharmonic voicing with fully
convolutional neural networks}

\author{
      Takeshi Ikuma \IEEEmembership{Member, IEEE}, Melda Kunduk, Brad
Story, and     Andrew J. McWhorter
\thanks{T. Ikuma is with Department of Otolaryngology--Head and Neck
Surgery, Louisiana State University Health Sciences Center, New Orleans,
LA.}
\thanks{M. Kunduk is with Department of Communication Disorders,
Louisiana State University, Baton Rouge, LA.}
\thanks{B. Story is with Dept. of Speech, Language, Hearing Sciences,
University of Arizona.}
\thanks{A. J. McWhorter is with Department of Otolaryngology--Head and
Neck Surgery, Louisiana State University Health Sciences Center, New
Orleans, LA.}
\thanks{This manuscript is an extended version of an abstract submitted
to the 16th International Conference on Advances in Quantitative
Laryngology, Voice and Speech Research, Groningen, the Netherlands, June
24th - 27th, 2025.}
}

\markboth{Submitted to IEEE Trans Audio Speech Lang Process, Januray
2025}%
{Detection of subharmonic voicing}

\maketitle

\begin{abstract}
Many voice disorders induce subharmonic phonation, but voice signal
analysis is currently lacking a technique to detect the presence of
subharmonics reliably. Distinguishing subharmonic phonation from normal
phonation is a challenging task as both are nearly periodic phenomena.
Subharmonic phonation adds cyclical variations to the normal glottal
cycles. Hence, the estimation of subharmonic period requires a wholistic
analysis of the signals. Deep learning is an effective solution to this
type of complex problem. This paper describes fully convolutional neural
networks which are trained with synthesized subharmonic voice signals to
classify the subharmonic periods. Synthetic evaluation shows over 98\%
classification accuracy, and assessment of sustained vowel recordings
demonstrates encouraging outcomes as well as the areas for future
improvements.
\end{abstract}

\begin{IEEEkeywords}
Disordered voice, Acoustic voice analysis, Deep learning
\end{IEEEkeywords}

\section{Introduction}\label{introduction}

Subharmonic phonation is a voicing mode in which the vibration pattern
of vocal folds varies from cycle to cycle before repeating itself. It is
often associated with vocal pathology, especially those which alter the
physical characteristics of the vocal folds such as vocal fold lesions
and vocal fold paralysis
\citep{behrman1998, cavalli1999, nunezbatalla2000, kramer2013, ikuma2023a}.
Subharmonic voicing is also possible for vocally healthy speakers.
Subharmonics can be voluntarily produced as vocal fry
\citep{hollien1966} or as a singing technique \citep{herbst2017}.
Perceptually, the subharmonic voice presents rough voice quality
\citep{omori1997} or lowered pitch
\citep{bergan2001, sun2002a, huang2024}.

The presence of subharmonics may degrade the accuracy of the voice
parameter measurements used in clinical acoustic analysis. This stems
from the primary property of the subharmonic signals, that is, they are
nearly periodic as is the normal phonation. This similarity often causes
existing fundamental frequency estimators to yield erroneous estimates
under strong subharmonics \citep{ikuma2024d, ikuma2025}. The (speaking)
fundamental frequency \(f_o\) that those algorithms estimates is the
foundation of the most of voice parameters and should represent the
cycle frequency of glottal opening and closing, disregarding the
variabilities in cyclic variations in amplitude, frequency, or shape.
Incorrectly using the subharmonic fundamental frequency (e.g, \(f_o/2\)
for period doubling subharmonics) as the \(f_o\) likely results in the
final parameter measures to under-represent the severity of pathological
voice. As such, an accurate numerical solution to detect the presence of
the subharmonics is important for the pathological voice analysis.

Titze \citep{titze1994} in 1994 categorized voice signals which
encompass the subharmonic behaviors as type 2 in his voice signal typing
system. His recommended approach to analyze type 2 signals is by
visualization tools and subjective inspection and discouraged the use of
numerical measures citing their unreliability. Since 1994, little
success has been reported to analyze subharmonic voice signals. The NSH
(number of subharmonics) parameter of KayPENTAX MDVP software
\citep{kaypentax2008, deliyski1993} reportedly has a poor accuracy
\citep{cavalli1999}. Sun \citep{sun2000} introduced the concept of the
subharmonic-to-harmonic ratio (SHR) and used it to estimate the
fundamental frequency of voice signals with a possible presence of
period-doubling subharmonics, and Hlavni\v{c}ka et al.
\citep{hlavnicka2019} strengthened Sun's method with improved initial
\(f_o\) estimation. Sun's approach, however, is not the ideal solution
to compute the SHR as there is a conundrum of trying to measure the SHR
without knowing the presence of the subharmonics. These methods are also
limited to detect period-doubling subharmonics and not higher-period
subharmonics that are also prevalent in pathological voice
\citep{ikuma2023a}. Aichinger et al. \citep{aichinger2017b} proposed to
use two harmonic models to analyze biphonic voice (i.e., two parts of
glottis vibrating at different frequencies) but a strong coupling of the
two sources, which induces subharmonics, causes the analysis to fail
\citep{aichinger2018a}. Finally, Awan and Awan \citep{awan2020} proposed
two-stage cepstrum peak prominence difference to measure vocal
roughness. While this parameter was derived to measure the difference
between the \(f_o\) and the subharmonic (or true) fundamental frequency,
it requires the approximate range of the expected \(f_o\), and it
reportedly failed to report subharmonics sufficiently
\citep{kitayama2023}.

Accurate detection of incidental subharmonics over a short segment of
voice signal leads to improve the subharmonic measures such as NSH and
SHR and to aid the \(f_o\) estimation, thereby also improving the
qualities of the vocal parameters at large. Detection of subharmonics is
synonymous to estimation of the subharmonic period \(M\), noting that
\(M=1\) naturally lend itself to the normal phonation. This paper
explores deep learning (DL) as a potential solution to classify the
subharmonic period. This idea is motivated by the DL fundamental
frequency estimators \citep{ardaillon2019, kim2018}. These estimators
outperform the conventional estimators and demonstrated their robustness
against the presence of subharmonics \citep{ikuma2024d}. If a DL network
can be trained to ignore subharmonics, a similar network could likely be
trained to find subharmonics. We specifically investigate the efficacy
of fully convolutional neural network (FCN) architecture.

The rest of paper is organized as follows. Section \ref{sec:fcn} defines
the DL networks under study and Section \ref{sec:synthesis} establishes
the subharmonic signal synthesis framework used to train and evaluate
the networks. The performance of the trained networks is presented with
the synthesized signals in Section \ref{sec:synth-res} and with real
sustained vowel recordings in Section \ref{sec:sustained-vowel-res},
followed by the concluding remarks.

\section{Fully convolutional neural networks}\label{sec:fcn}

Clinical voice recordings have arbitrary durations, and this makes the
fully convolutional neural network (FCN) architecture
\citep{long2015, wang2017, ismailfawaz2019} a suitable DL model as is
used for the \(f_o\) estimation \citep{ardaillon2019}. Time-varying
nature of voice signals requires most of their analysis parameters to be
computed over a short sliding window, and the subharmonic period is no
different as subharmonic voicing in pathological voice can occur
intermittently and possibly in a rapid succession
\citep{behrman1998, ikuma2023a, titze1994, dejonckere1983}. FCN with
local pooling layers is fit to perform this task efficiently with
overlapping snapshots as a pooling layer acts as a generalized
downsampler, thus decoupling the window size and the snapshot interval.

The input sampling rate \(f_s\) is set to 8000 samples per second (S/s),
which is sufficient to estimate the vocal fundamental frequency
(\(f_o\)) \citep{ardaillon2019}, and thus also sufficient for
subharmonic period. Two FCNs are evaluated with different target window
sizes: 400 samples (50 ms) and 800 samples (100 ms). The shorter window
is designed to be more robust against intermittency while the longer
window aims to improve the classification performance of sustained
subharmonics. The output interval is set to 2 ms or 500 samples/second,
requiring the pooling layers collectively to reduce the sampling rate by
8.

The subharmonic period \(M\) can theoretically be any positive integer;
however, the vast majority is period doubling (\(M=2\)) and the
prevalence rapidly diminishes as \(M\) increases
\citep{ikuma2023a, ikuma2024d, hlavnicka2019}. Also, a signal snapshot
must contain more than \(M+1\) glottal cycles for the classifier to
determine the subharmonic period \(M\) with any confidence. The maximum
\(M\) is set to four with the label set \(\mathcal{M}=\{1,2,3,4\}\),
based on the target window sizes and expected range of \(f_o\) (100-300
Hz).

\begin{figure*}
\centering
\includegraphics[keepaspectratio]{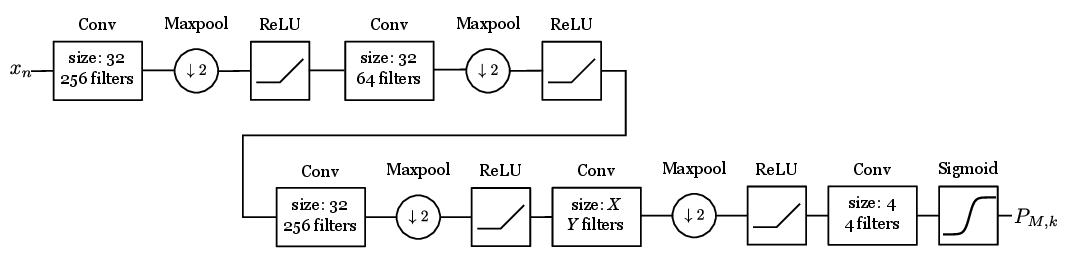}
\caption{Fully convolutional neural network architecture under study:
FCN-401 with \(X=16\) and \(Y=512\), and FCN-785 with \(X=64\) and
\(Y=128\). All convolutional layers are applied with a stride of 1, and
there is a batch normalization layer (not pictured) before every
ReLU.}\label{fig:fcn}
\end{figure*}

The FCN structure under study comprises 5 convolutional layers as
depicted in Fig. \ref{fig:fcn}. The input signal \(x_n\) is a
pre-normalized acoustic signal of arbitrary duration, and the first 4
convolutional layers are accompanied by max pooling two consecutive
samples, each effectively downsampling the propagating signal by 2 and
collectively by 8. These convolutional layers are also followed by batch
normalization and rectified linear unit (ReLU) activation function. The
final output convolutional layer has four channels to sigmoid activation
functions. The network outputs
\(\left\{ P_{M,k}: M \in \mathcal{M} \right\}\), the probabilities of
the \(k\)th snapshot contains the signal with subharmonic period \(M\).
The softmax activation is not used at the output because the subharmonic
period candidates in \(\mathcal{M}\) is not exhaustive set of
pathological voice signals. The number of filters and their sizes of
each convolution layer are denoted in Fig. \ref{fig:fcn}. The same
network architecture is used for both of the target window sizes by
using different allocations of its coefficients for the fourth
convolution layer. The first network, named FCN-401, with 401-sample
(50.1-ms) window size uses many short filters (512 filters of
16-coefficient filters) while the second network, FCN-785 with
785-sample (98.1-ms) window size with a few long filter (128 length-64
filters). Thus, both networks have the same number of coefficients and
incur nearly the same computational complexity.

\section{Synthesis of training dataset}\label{sec:synthesis}

For the DL training and evaluation, having a dataset with known truths
is essential. Clinical voice recordings require manual annotation, and
the accuracy of manual annotation is questionable because of the nearly
periodic nature of subharmonic signals. Hence, we turned to a numerical
voice synthesis model to parameterize the subharmonic voicing and
generate a large dataset via Monte Carlo simulation.

To study the subharmonic phonation, prior work in literature also
utilized synthesis tools to generate voice signals with subharmonics
\citep{bergan2001, hlavnicka2019, herbst2021, huang2024}. These studies
were all based on introducing amplitude or frequency modulation to the
source glottal flow signal via cycle-wise manipulation of the signal and
were limited to period-doubling subharmonics. This general approach
limits the complexity of the generated signals because the flow pulse
shapes are merely time- or amplitude-scaled versions.

In this study, the synthesis process employed the kinematic vocal fold
model \citep{titze1984, titze1989}, which was aerodynamically and
acoustically coupled to two-port wave propagation vocal tract model
\citep{story1995a, liljencrants1985}. This synthesis framework was
previously used to study breathy voice \citep{samlan2011, samlan2013}
and vocal fold asymmetry \citep{samlan2014a}. The kinematic model was
also used to simulate the period-doubling electroglottographic (EGG)
signals although the subharmonics were introduced post-synthesis
\citep{herbst2021}. In the current work, the subharmonic modulation is
introduced to the vocal fold vibration so that the modulation is subject
to the vocal fold collision and the nonlinear relationship between the
glottal area and glottal flow. This approach realizes a more diverse set
of subharmonics than previous synthesis approaches and naturally extends
to the generation of higher-period subharmonics.

The kinematic vocal fold model \citep{titze1984, titze1989} is built
around a three-dimensional geometry of symmetric vocal folds. In
exchange for removing the displacement-force coupling of mass-spring
models \citep{ishizaka1972, titze1988}, the kinematic model concretely
defines the movements of the vocal folds. This enables simple and direct
incorporation of subharmonic vibration to the model. The displacement
function without the presence of the other vocal fold is modeled by
\begin{equation} \label{eq:tilde-xi}
\tilde{\xi}(y,z,t) = \xi_0(y,z) + \xi_m \sin (\pi y /L) r(t;\varphi),
\end{equation}
where \(L\) is the vibrating glottal length, \(T\) is the vibrating
glottal thickness, \(\xi_0(y,z)\) is the prephonatory edge position,
\(\xi_m\) is the maximum vocal fold displacement, \(r(t;\varphi)\) is
the reference vibration function, and
\begin{equation}{\varphi = -2 \pi Q_p (z/T - R_{zn}),
}\end{equation} is the phase delay between the upper and lower edges of
the folds. Here. \(Q_p\) is the phase quotient \citep{titze1984} and
\(R_{zn}\) is the nodal point ratio which specifies the pivot point of
vocal fold rotation relative to the vocal fold thickness \(T\)
\citep{samlan2011}. Since symmetrical, the vocal folds collide on the
glottal midline, resulting in the displacement function:
\begin{equation}{\xi(y,z,t) = \begin{cases}\tilde{\xi}(y,z,t) & \text{if $\tilde{\xi}(y,z,t)>0$,} \\
0 & \text{else.}\end{cases}
}\end{equation} The glottal opening is then given by the minimum
displacement along the depth,
\begin{equation}{\xi_\text{min}(y,t) = \min_z \xi(y,z,t),}\end{equation}
which leads to the glottal area: \begin{equation} \label{eq:ag}
a(t) = 2 \int_0^L \xi_\text{min}(y,t) dy.\end{equation}

The subharmonics are introduced to the glottal vibration via
\(r(t;\varphi)\). There are two types of subharmonic glottal vibrations
or a combination thereof. First is the modulation
\citep{kiritani1993, kniesburges2016, ikuma2016a}, where the entire
glottis exhibit subharmonic oscillatory behavior. Second is entrained
biphonation \citep{kiritani1993, mergell2000, neubauer2001, ikuma2016a},
where two parts of the glottis (e.g., left and right vocal fold, or
anterior and posterior regions) vibrate with different fundamental
frequencies but their ratio is rational. The dataset in this study only
comprises the subharmonic modulation with symmetrical vocal folds,
leaving the inclusion of subharmonic biphonations for future studies.
The modulation effect is implemented by a combination of amplitude
modulation (AM) and frequency modulation (FM), and subharmonic reference
vibration with period \(M \in \mathbb{I}\), \(M>1\), takes the form:
\begin{equation}{\begin{aligned}
r_M&(t;\varphi) = \left[ 1 + \epsilon_\text{AM} \sin \left(\frac{\phi(t;\varphi)}{M} + \phi_\text{AM}\right)\right]\\
 &\times \sin \left[ \phi(t;\varphi) + \epsilon_\text{FM} M \sin \left( \frac{\phi(t;\varphi)}{M}  + \phi_\text{FM}\right) \right], \\
\end{aligned}
}\end{equation} where
\begin{equation}{\phi(t;\varphi) = 2 \pi f_o t - \varphi,}\end{equation}
\(f_o\) is the speaking fundamental frequency, \(\epsilon_\text{AM}\) is
the AM extent, \(\phi_\text{AM}\) is the AM phase,
\(\epsilon_\text{FM}\) is the FM extent, and \(\phi_\text{FM}\) is the
FM phase. Including the normal mode, we have
\begin{equation}{r(t;\varphi) = \begin{cases} \sin \phi(t;\varphi) & \text{if } M=1,\\ 
                             r_M(t;\varphi) & \text{if } M>1.\end{cases}
}\end{equation} Here, \(\sin \phi(t;\varphi)\) is the reference
vibration used in the original model \citep{titze1984}. The AM and FM
extents generalize those of the period-doubling case
\citep{titze1994, herbst2021}:
\begin{equation}{\epsilon_\text{AM} = \frac{A_\text{max}-A_\text{min}}{A_\text{max}+A_\text{min}}
}\end{equation} and
\begin{equation}{\epsilon_\text{FM} = \frac{T_\text{max}-T_\text{min}}{T_\text{max}+T_\text{min}},
}\end{equation} where \(A_\text{max}\) and \(A_\text{min}\) are the
extrema of the local maxima over both \(t\) and \(\phi_\text{AM}\) while
\(T_\text{max}\) and \(T_\text{min}\) are the extrema of cycle periods
over both \(t\) and \(\phi_\text{FM}\).

The prephonatory position \(\xi_0\) of the vocal folds is defined by
\citep{titze1989}
\begin{equation}{\xi_0(y,z) = \left[ Q_a + \left( Q_s - 4 Q_b \frac{z}{T}\right)\left( 1 - \frac{z}{T}\right) \right]\left( 1 - \frac{y}{L}\right),
}\end{equation} where \(Q_a\) is the abduction coefficient, \(Q_s\) is
the shape coefficient, and \(Q_b\) is the bulging quotient.

\begin{figure}
\centering
\includegraphics[keepaspectratio]{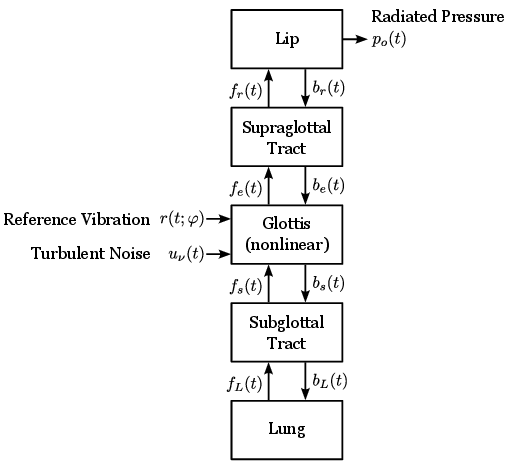}
\caption{Block diagram of the transmission-line voice synthesis
model.}\label{fig:synth-blockdgm}
\end{figure}

The glottal area vibration in Eq. \eqref{eq:ag} is coupled to the
propagations of incidental pressure waves as depicted in
Fig. \ref{fig:synth-blockdgm}. The model consists of five blocks: lung,
subglottal tract, glottis, supraglottal tract, and lip radiation. Each
block receives a forward incidental pressure wave \(f_1(t)\) from below
and a backward wave \(b_2(t)\) from above and produces a forward wave
\(f_2(t)\) to above and backward wave \(b_1(t)\) to below.

The subglottal and supraglottal vocal tracts are crudely modeled as
leaky uniform tubes with crosssectional areas \(A_s\) and \(A_e\),
respectively. Given a tract length \(L_T\), propagated incidental
pressure waves are given by \begin{equation}{\begin{aligned}
f_2(t) &= \alpha^{L_T} f_1(t - cL_T) \text{ and} \\ 
b_1(t) &= \alpha^{L_T} b_2(t - cL_T), \\
\end{aligned}
}\end{equation} where \(\alpha<1\) is the propagation gain per unit
length, and \(c\) is the speed of sound.

At the glottis, the glottal flow \(u(t)\) is dictated by the glottal
area \(a(t)\) in Eq. \eqref{eq:ag} and the subglottal forward pressure
\(f_s(t)\) and supraglottal backward pressure \(b_e(t)\)
\citep{titze1984}: \begin{equation} \label{eq:u}
\begin{aligned}
u(t) &= \frac{a(t)c}{1-k_e(t)} \Bigg\{ -\frac{a(t)}{A^*}\\
&\pm \left[\left(\frac{a(t)}{A^*}\right)^2 + \frac{4(1-k_e(t))}{c^2\rho}\left(f_s(t)-b_e(t)\right)\right]^\frac{1}{2}
\Bigg\},
\end{aligned}
\end{equation}
where \(A^* = (A_e^{-1}+A_s^{-1})^{-1}\) is the effective vocal tract
area, \(\rho\) is the density of the air, and the pressure recovery
coefficient \citep{titze2002},
\begin{equation}{k_e(t) = 2\frac{a(t)}{A_e}\left(1-\frac{a(t)}{A_e}\right).
}\end{equation} In Eq. \eqref{eq:u}, the plus sign is used when the
second term in the square brackets is negative, and vice versa.

In addition to \(r(t;\varphi)\), the turbulent noise \(u_\nu(t)\) is
another input to glottis. The noise generation follows a mixture of the
noise source of the Klatt voice simultor \citep{klatt1990} and the noise
level switching mechanism based on Reynolds number \citep{samlan2011}.
It intends to model the aspiration noise during high flow rate (i.e.,
most open phase of vocal fold vibration) as well as the low-level noise
caused by dc flow \citep{holmberg1988}: \begin{equation} \label{eq:u_nu}
u_\nu(t) = \begin{cases}
\nu(t) & \text{for $Re>Re_c$}\\
\delta \nu(t) & \text{for $Re \le Re_c$,}\\
\end{cases}
\end{equation}
where \(\nu(t)\) is the common noise source,
\begin{equation}{Re = \frac{u(t)\rho}{L\mu}
}\end{equation} is the Reynolds number, \(Re_c=1200\) is the critical
Reynolds number, and \(\delta\) is a coefficient to supress the noise
level in the subcritical region. The noise \(\nu(t)\) is modeled as a
first-order lowpass Gaussian noise with \(-20\) dB rolloff and dc power
spectral density level \(S_\nu(0)\). Combining Eq. \eqref{eq:u} and
Eq. \eqref{eq:u_nu}, the total glottal flow is given by
\begin{equation}{u_g(t) = u(t) + u_\nu(t),
}\end{equation} and the incidental pressure wave outputs:
\begin{equation}{b_e(t) = f_e(t) - \frac{\rho c}{A_e} u_g(t),
}\end{equation} and
\begin{equation}{f_s(t) = b_s(t) + \frac{\rho c}{A_s} u_g(t).
}\end{equation}

The lung drives the model with a constant pressure \(P_L\), and its
boundary condition is modeled by
\begin{equation}{f_L(t) = 0.9 P_L - 0.8 b_L(t).
}\end{equation} This approximately matches the lung and subglottal tract
impedances. The lip radition \(p_o(t)\) is modeled by a circular piston
in an infinite baffle \citep{story1995a, ishizaka1972}:
\begin{equation}{\left[\begin{array}{c}P_o(s)\\ B_r(s)\end{array}\right] = \left[\begin{array}{c}2 Z_r(s) \\ Z_r(s) - 1\end{array}\right] \left[Z_r(s) + 1\right]^{-1}F_r(s),
}\end{equation} where \(P_o(s)\), \(B_r(s)\), and \(F_r(s)\) are the
Laplace transforms of \(p_o(t)\), \(b_r(t)\), and \(f_r(t)\),
respectively, and \begin{equation}{Z_r(s) = \frac{sL_r R_r}{R_r + sL_r}
}\end{equation} is the lip impedance with
\(L_r = 8/(3\pi c) \sqrt{A_e/\pi}\) and \(R_r = 128 / (9\pi^2)\).

To run the synthesis, the model was discretized both in time and space.
The vocal fold surface is sampled to have 21 cells along the \(y\) axis
and 15 cells along the \(z\) axis. The sampling rate of the simulation
is \(f_s=\) 44100 S/s, which also quantizes the possible vocal tract
lengths to integer multiples of \(c/f_s=\) 0.794 cm. The lip radiation
model is converted to a discrete-time model by the bilinear
transformation.

All the free parameters of the synthesis are listed in
Table \ref{tbl:synth-params}. Except for \(M\), all 21 other parameters
were randomly generated during the Monte Carlo simulation. They were
drawn from independent uniform distributions with the value ranges as
specified on the table. Each synthesized data was generated for 1.1
seconds with fixed parameter values, thereby simulating sustained
vowel-like signal. The generated signal was then resampled down to 8000
S/s, and the first 0.1 seconds (800 samples) were purged to keep only
the steady-state portion of the signal. Finally, the signal was
normalized by its mean and variance.

\begin{table*}[h!]
\caption{Synthesis Parameters}
\label{tbl:synth-params}
\begin{tabular}{lllll}
Block/Signal&Parameter&Range&Units&Description\\
\hline
$r(t;\varphi)$&$M$&\{1, 2, 3, 4\}&&Subharmonic period\\
&$f_o$&[100, 300),&Hz&Speaking fundamental frequency\\
&$\epsilon_\text{AM}$&[0.1, 1.0)&&AM extent (log-uniform)\\
&$\epsilon_\text{FM}$&[0.005, 0.1)&&FM extent (log-uniform)\\
&$\phi_\text{AM}$&$[-\pi/2,\pi/2)$&rad&AM phase\\
&$\phi_\text{FM}$&$[-\pi/2,\pi/2)$&rad&FM phase\\
$u_\nu(t)$&$S_\nu(0)$&[100, 2500)&(cm³/s)²/Hz&DC psd level of full aspiration noise\\
&$\delta$&[0.2, 0.6)&&Power reduction factor during subcritical phase\\
Glottis&$L$&[0.738, 1.562)&cm&Vibrating vocal fold length\\
&$T$&[0.18, 0.33)&cm&Vibrating vocal fold thickness\\
&$\xi_m$&[0.09, 0.132)&cm&Maximum glottal half width\\
&$Q_a$&[0.27, 0.33)&&Abduction quotient\\
&$Q_s$&[1.8, 2.2)&&Shape quotient\\
&$Q_b$&[$0.45 Q_s$, $0.55Q_s$)&&Bulging quotient (relative to $Q_s$)\\
&$Q_p$&[0.18, 0.22)&&Phase quotient\\
&$R_{zn}$&[0.63, 0.77)&&Nodal point ratio\\
Vocal Tracts&$\alpha$&[0.9983, 0.9985)&1/cm&Common propagation gain per length\\
Supraglottal Tract&$L_T$&[11.111, 15.873)&cm&Supraglottal tract length\\
&$A_e$&[1.0, 5.0)&cm²&Supraglottal tract area\\
Subglottal Tract&$L_T$&[6.349, 9.524)&cm&Subglottal tract length\\
&$A_s$&[1.0, 3.0)&cm²&Subglottal tract area\\
Lung&$P_L$&[7056, 8624)&dyn/cm²&Lung pressure\\
\end{tabular}
\end{table*}

\section{Training and evaluation with synthesized
dataset}\label{sec:synth-res}

To train the FCNs, 8000 1-second signals for each subharmonic period
target \(M \in \mathcal{M}\) were generated for training (32000 total)
and another 2000 each for validation (8000 total). Both models are
trained using gradient descent with the Adam optimizer
\citep{kingma2015} with mini-batches composed of 32 signals selected
from the training dataset in a random order. The learning rate of 0.0002
was used. An input dropout layer was inserted during trainig with 20\%
dropout rate. The training persisted over 20 epochs and the coefficients
with the best validation accuracy was chosen. All the snapshots of every
signal were used in the training to capture all the possible phases of
the periodic signal. Due to the different window sizes, FCN-401 and
FCN-785 have different numbers of snapshots per signal: 478 for FCN-401
and 454 for FCN-785.

To evaluate the trained networks, another 4000 signals were generated.
For these, \(f_o\) and the SHR were recorded along with \(M\). The SHR
was estimated from the periodogram \(S(f)\) of the signal with Hamming
window by
\begin{equation}{SHR = \frac{\sum\limits_{k \in \mathcal{K}_s} S\left(k f_o/M\right)}{\sum\limits_{k\in \mathcal{K}_h} S\left(k f_o/M\right)},
}\end{equation} where \(\mathcal{K}_h\) is a set of the \(f_o\)-harmonic
multipliers under 4 kHz, and \(\mathcal{K}_s\) is a set of the
subharmonic multipliers. Given the \(k\)th output of an FCN,
\(\{P_{M,k} : M \in \mathcal{M}\}\), its subharmonic period is estimated
by \(\hat{M}_k = \operatorname{argmax}_{M \in \mathcal{M}} {P_{M,k}}\).

\begin{figure}
\centering
\includegraphics[keepaspectratio]{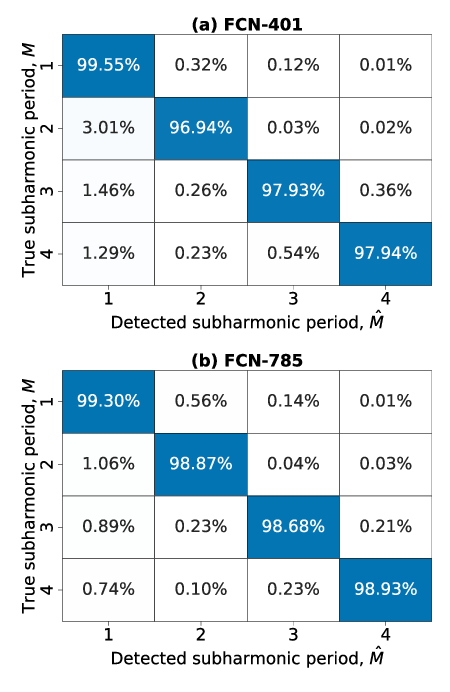}
\caption{Sythetic classification confusion matrices: (a) FCN-401, (b)
FCN-785.}\label{fig:confmtx}
\end{figure}

The overall classification accuracy of FCN-401 is 98.1\% and that of
FCN-785 is 98.9\%. The confusion matrices of the two FCNs are shown in
Fig. \ref{fig:confmtx}. FCN-785 achieved above 98.5\% conditional
accuracies across all \(M\) with 0.5\% higher results for the normal
voicing detection (\(M=1\)). Subharmonic voicings (\(M>1\)) are more
likely to be mistaken as normal (\(\hat{M}=1\)) than with another
subharmonic period (\(\hat{M}>1\)). FCN-401 follows the same trend but
with increased errors. Its normal voicing detection matches that of
FCN-785, but the subharmonic period detections lags behind by about
0.75\% or more.

\begin{figure}
\centering
\includegraphics[keepaspectratio]{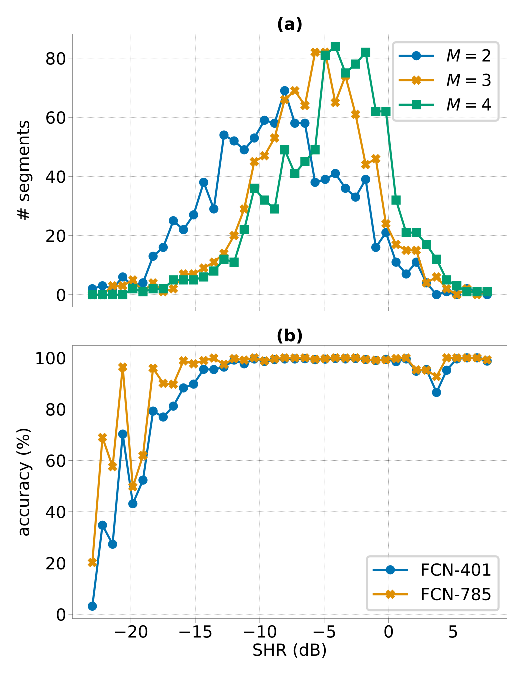}
\caption{Synthetic classification performance vs.~SHR (\(M>1\) signals
only): (a) SHR distribution and (b) classification accuracy
vs.~SHR.}\label{fig:shr}
\end{figure}

FCN-401 performed noticeably worse for the \(M=2\) signals (96.94\%
acuracy). In contrast, FCN-785 maintained a similar accuracy across all
subharmonic periods. The observed performance loss by FCN-401 for
\(M=2\) likely stems from two reasons: SHR imbalance across \(M\) and
short window size. The SHRs of the acoustic signals are not identical as
shown in Fig. \ref{fig:shr}(a) although the identical distributions were
used for AM and FM extents across all \(M\). The expected SHR of \(M=2\)
is lower than \(M>2\) signals, and the presence of the \(M>2\) signals
with SHR below \(-10\) dB rapidly declines. This likely deemphasized the
importance of the low SHR range during the training, resulting in poor
accuracy as shown in Fig. \ref{fig:shr}(b). A compounding factor is that
lower the SHR the harder to detect the subharmonics.
Fig. \ref{fig:shr}(b) also displays that the subharmonic detection
performs better with a longer analysis window.

\begin{figure}
\centering
\includegraphics[keepaspectratio]{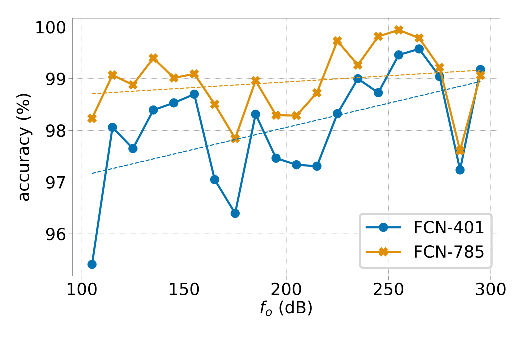}
\caption{Synthetic classification performance vs.~\(f_o\). Dashed lines
are the least squares fitted lines.}\label{fig:fo}
\end{figure}

Another effect which separated FCN-785 from FCN-401 is \(f_o\) as shown
in Fig. \ref{fig:fo}. There is a clear performance gap between the two
FCNs at lower \(f_o\) range. Although FCNs do not strictly follow the
rules of tranditional spectral analysis, a form of spectral resolution
seems to be playing a role here. In spectral analysis, resolving two
tones which are separated by 50 Hz (i.e., tones of the \(M=2\) signals
at 100 Hz) requires a snapshot size of at least 40 ms (320 samples)
under noise-free condition. An even longer snapshot is needed for the
low-SHR signals as the subharmonic tones are much weaker than the
harmonic tones. FCN-401 is operating close to this limit; thus, its
accuracy suffers as \(f_o\) reduces. FCN-785, on the other hand, can
maintain its accuracy across \(f_o\).

\section{Case studies of pathological sustained
vowels}\label{sec:sustained-vowel-res}

Next, the behaviors of the trained FCNs are evaluated with sustained /a/
vowel recordings of selected pathological voices from KayPENTAX
(Massachusetts Eye and Ear Infirmary) Disordered Voice Database
\citep{kaypentax2006}. These recordings are sampled at 25000 S/s and
each about 1 second long. The onsets and offsets of the phonations are
pre-trimmed. To evaluate, the recordings are first resampled to 8000 S/s
and normalized over the entire duration before input to the FCNs. The
estimated subharmonic period sequence,
\(\hat{M}_k = \operatorname{argmax}_{M\in\mathcal{M}} P_{M,k}\), and the
probabilities of the estimates \(P_{\hat{M}_k,k}\) are observed for both
FCN-401 and FCN-785.

\begin{figure}
\centering
\includegraphics[keepaspectratio]{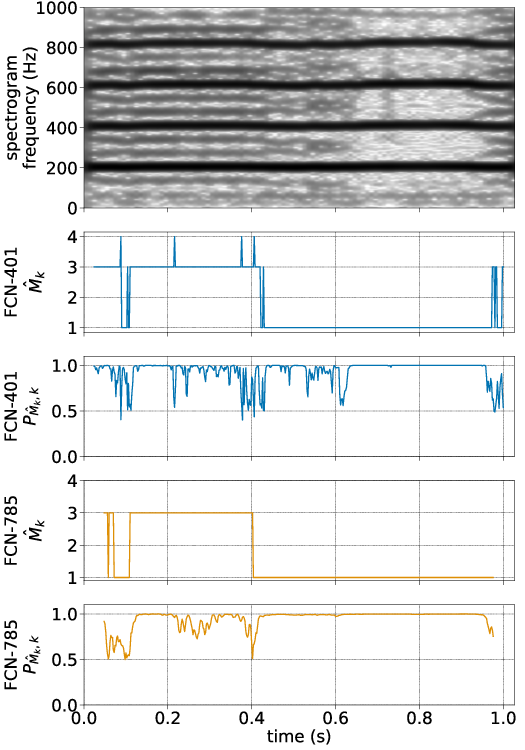}
\caption{Case Study 1 (modulation): Spectrogram and outputs of trained
FCNs---estimated subharmonic periods \(\hat{M}_K\) and their
probabilities \(P_{\hat{M},k}\).}\label{fig:meei1}
\end{figure}

The first case in Fig. \ref{fig:meei1} consists of \(f_o\approx206\) Hz
phonation, starting with period-tripling subharmonics until \(t=0.42\)
s, followed by a brief period of unlocked modulation (\textless0.63 s)
to normal phonation (\textless0.92 s) before the period-3 returning.
Both networks estimate either \(M=1\) or \(3\), except for FCN-401
occasionally reporting low-probability \(M_k=4\) snapshots. This case
exemplifies the strictness of the FCNs in detecting the subharmonics.
Despite unlocked modulation was not included in the training dataset,
both selected \(M=1\) rather than visually viable \(M=2\). Also,
consistent near-1.0 probabilities of \(M=3\) cases were obtained only
between 0.1 and 0.2 s, possibly indicating a weaker lock. Finally, this
case demonstrates the tradeoff of FCN-785. While its longer window size
provides better estimation performance than FCN-401, it also negatively
impact the ability to detect the transition point. At \(t=0.4\), it
leaves \(M_k=3\) slightly earlier than when it occurs in the spectrogram
and FCN-401 results.

\begin{figure}
\centering
\includegraphics[keepaspectratio]{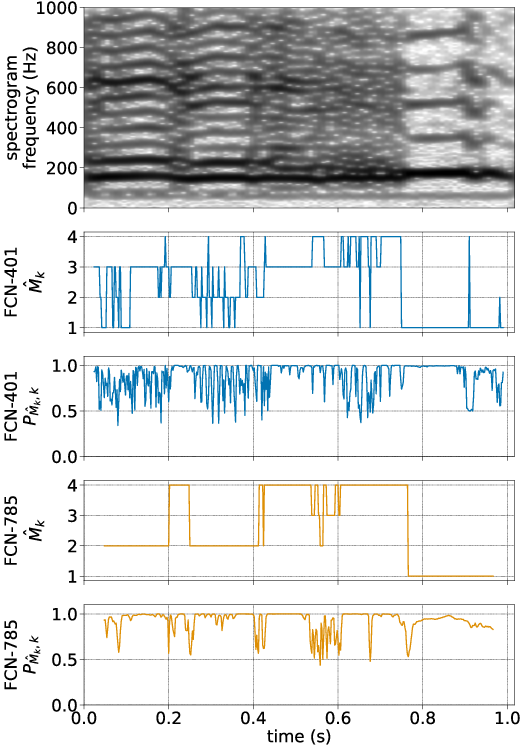}
\caption{Case Study 2 (suspected biphonia): Spectrogram and outputs of
trained FCNs---estimated subharmonic periods \(\hat{M}_K\) and their
probabilities \(P_{\hat{M},k}\).}\label{fig:meei2}
\end{figure}

The second case in Fig. \ref{fig:meei2} demonstrates the outputs of the
trained FCNs when the input signal contains suspected biphonation, which
is apparent until \(t=0.75\) s. The biphonation is suspected because the
traces of two dominant low-frequency tones are consistently present,
each likely representing the first harmonic of its periodic vibration.
At the beginning, \(f_{o,1}= 154\) Hz and \(f_{o,2}= 231\) Hz, which
consititute a \(2:3\) subharmonic arrangement. Then, biphonation
continues to exist through what appears to be less stable segment
between 0.44 and 0.75 s. While we can continue to observe the two first
harmonics, whether they are locked or not is no longer prevalent to
naked eyes. Although the FCNs are not trained for this type of
subharmonics, both FCNs suggest that there are moments of locked
oscillation with high confidence. FCN-785 often estimates
\(\hat{M}_k=4\) over sustained periods in the unstable segment.
Together, it appears that we have 3:4 in the first half and 4:5 in the
second half. This is an encouraging results, demonstrating the
robustness of the FCN-based subharmonic detection. The subharmonics due
to biphonation has a different spectral construct than those due to the
modulation, but high confident estimates are obtained despite the
training dataset only contains modulation signals only. FCNs would
likely perform better if the subharmonic biphonation cases are included
in the training data.

\begin{figure}
\centering
\includegraphics[keepaspectratio]{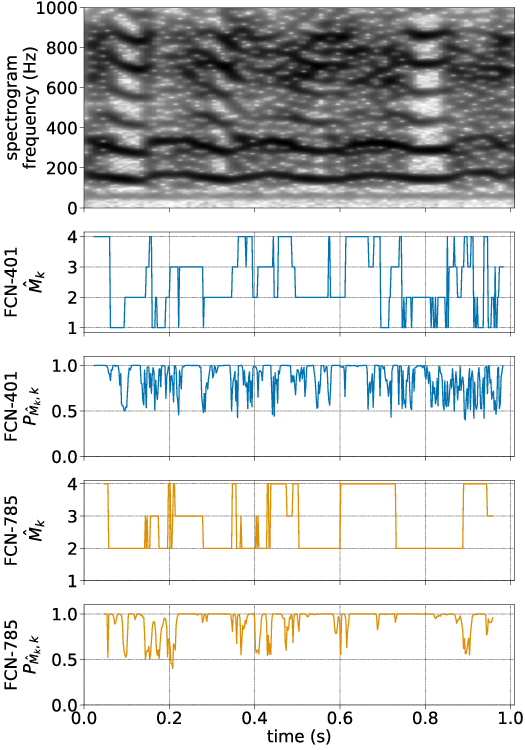}
\caption{Case Study 3 (severe vocal tremor): Spectrogram and outputs of
trained FCNs---estimated subharmonic periods \(\hat{M}_K\) and their
probabilities \(P_{\hat{M},k}\).}\label{fig:meei3}
\end{figure}

The final case in Fig. \ref{fig:meei3} illustrates the behaviors of the
trained FCNs for a signal with severe vocal tremor, which includes
\(f_o\) tremor with the period of 4 s and peak-to-peak extent of 35 Hz.
Neither FCNs consistently estimated \(\hat{M}_k=1\) during the short
segments of clean harmonics (\(t\): 0.05-0.13 and 0.77-0.84). FCN-401's
inability to maintain \(\hat{M}_k=1\) for the entirity of these segments
is likely due to its training dataset only consists of fixed \(f_o\).
FCN-785, which does not register \(\hat{M}_k=1\) at all, is further
penalized by the shortness of these segments coupled with the
homogeneous training dataset. This indicates that the training data must
include \(f_o\) variability (e.g., the flutter effect of the Klatt
synthesizer \citep{klatt1990}) to improve the performance. This is also
crucial to extend the use to connected speech, in which \(f_o\)
constantly fluctuates.

\section{Conclusions}\label{conclusions}

The synthetic evaluation indicates that the proposed FCNs are highly
capable of correctly identifying the subharmonic periods. Testing with
pathological voice recordings further supports indicates that the FCNs'
potential to identify subharmonic modulations despite being trained with
a crude vocal tract model and arbitrarily selected parameter ranges.
This also demonstrates the FCNs' ability to ignore weak unlocked
modulation. The testing also revealed a number of remaining challenges
to be fulfilled. First, the pathological voice simulation needs to be
extended to account for additional pathological behaviors like
subharmonic biphonation, intermittency, and variation in fundamental
frequency. Second, a better stochastical representation of human voice
production system is crucial to refine the synthesis parameter ranges.
These will yield improved training dataset, and subsequent tuning of the
FCN architecture and hyperparameters will likely result in a dependable
subharmonic detector.



\begin{thebibliography}{48}
\providecommand{\natexlab}[1]{#1}
\providecommand{\url}[1]{#1}
\csname url@samestyle\endcsname
\providecommand{\newblock}{\relax}
\providecommand{\bibinfo}[2]{#2}
\providecommand{\BIBentrySTDinterwordspacing}{\spaceskip=0pt\relax}
\providecommand{\BIBentryALTinterwordstretchfactor}{4}
\providecommand{\BIBentryALTinterwordspacing}{\spaceskip=\fontdimen2\font plus
\BIBentryALTinterwordstretchfactor\fontdimen3\font minus
  \fontdimen4\font\relax}
\providecommand{\BIBforeignlanguage}[2]{{%
\expandafter\ifx\csname l@#1\endcsname\relax
\typeout{** WARNING: IEEEtranN.bst: No hyphenation pattern has been}%
\typeout{** loaded for the language `#1'. Using the pattern for}%
\typeout{** the default language instead.}%
\else
\language=\csname l@#1\endcsname
\fi
#2}}
\providecommand{\BIBdecl}{\relax}
\BIBdecl

\bibitem[Behrman et~al.(1998)Behrman, Agresti, Blumstein, and Lee]{behrman1998}
A.~Behrman, C.~J. Agresti, E.~Blumstein, and N.~Lee,
  ``\href{https://doi.org/10.1016/S0892-1997(98)80045-3}{Microphone and
  electroglottographic data from dysphonic patients: {{Type}} 1, 2 and 3
  signals},'' \emph{J. Voice}, vol.~12, no.~2, pp. 249--260, Jan. 1998.

\bibitem[Cavalli and Hirson(1999)]{cavalli1999}
L.~Cavalli and A.~Hirson,
  ``\href{https://doi.org/10.1016/S0892-1997(99)80009-5}{Diplophonia
  reappraised},'' \emph{J. Voice}, vol.~13, no.~4, pp. 542--556, 1999.

\bibitem[N{\'u}{\~n}ez~Batalla et~al.(2000)N{\'u}{\~n}ez~Batalla,
  Su{\'a}rez~Nieto, Mu{\~n}oz~Pinto, Baraga{\~n}o~R{\'i}o, Alvarez~Zapico, and
  Mart{\'i}nez~Ferreras]{nunezbatalla2000}
F.~N{\'u}{\~n}ez~Batalla, C.~Su{\'a}rez~Nieto, C.~Mu{\~n}oz~Pinto,
  L.~Baraga{\~n}o~R{\'i}o, M.~J. Alvarez~Zapico, and A.~Mart{\'i}nez~Ferreras,
  ``{[Spectrographic study of voice disorders: subharmonics]},'' \emph{Acta
  Otorrinolaringol. Esp.}, vol.~51, no.~1, pp. 52--56, 2000.

\bibitem[Kramer et~al.(2013)Kramer, Linder, and Sch{\"o}nweiler]{kramer2013}
E.~Kramer, R.~Linder, and R.~Sch{\"o}nweiler,
  ``\href{https://doi.org/10.1016/j.jvoice.2012.08.005}{A study of subharmonics
  in connected speech material},'' \emph{J. Voice}, vol.~27, no.~1, pp. 29--38,
  Jan. 2013.

\bibitem[Ikuma et~al.(2023)Ikuma, McWhorter, Adkins, and Kunduk]{ikuma2023a}
T.~Ikuma, A.~J. McWhorter, L.~Adkins, and M.~Kunduk,
  ``\href{https://doi.org/10.1044/2022_JSLHR-21-00499}{Investigation of vocal
  bifurcations and voice patterns induced by asymmetry of pathological vocal
  folds},'' \emph{J. Speech Lang. Hear. Res.}, vol.~66, no.~1, pp. 48--60, Jan.
  2023.

\bibitem[Hollien et~al.(1966)Hollien, Moore, Wendahl, and Michel]{hollien1966}
H.~Hollien, P.~Moore, R.~W. Wendahl, and J.~F. Michel,
  ``\href{https://doi.org/10.1044/jshr.0902.245}{On the nature of vocal fry},''
  \emph{J. Speech Hear. Res.}, vol.~9, no.~2, pp. 245--247, Jun. 1966.

\bibitem[Herbst et~al.(2017)Herbst, Hertegard, {Zangger-Borch}, and
  Lindestad]{herbst2017}
C.~T. Herbst, S.~Hertegard, D.~{Zangger-Borch}, and P.-{\AA}. Lindestad,
  ``\href{https://doi.org/10.3109/14015439.2016.1156737}{Freddie
  {{Mercury}}---acoustic analysis of speaking fundamental frequency, vibrato,
  and subharmonics},'' \emph{Logoped. Phoniatr. Vocol.}, vol.~42, no.~1, pp.
  29--38, Jan. 2017.

\bibitem[Omori et~al.(1997)Omori, Kojima, Kakani, Slavit, and
  Blaugrund]{omori1997}
K.~Omori, H.~Kojima, R.~Kakani, D.~H. Slavit, and S.~M. Blaugrund,
  ``\href{https://doi.org/10.1016/S0892-1997(97)80022-7}{Acoustic
  characteristics of rough voice: {{Subharmonics}}},'' \emph{J. Voice},
  vol.~11, no.~1, pp. 40--47, Mar. 1997.

\bibitem[Bergan and Titze(2001)]{bergan2001}
C.~C. Bergan and I.~R. Titze,
  ``\href{https://doi.org/10.1016/S0892-1997(01)00018-2}{Perception of pitch
  and roughness in vocal signals with subharmonics},'' \emph{J. Voice},
  vol.~15, no.~2, pp. 165--175, Jun. 2001.

\bibitem[Sun and Xu(2002)]{sun2002a}
X.~Sun and Y.~Xu,
  ``\href{https://doi.org/10.1016/S0892-1997(02)00119-4}{Perceived pitch of
  synthesized voice with alternate cycles},'' \emph{J. Voice}, vol.~16, no.~4,
  pp. 443--459, Dec. 2002.

\bibitem[Huang(2024)]{huang2024}
Y.~Huang, ``\href{https://doi.org/10.1121/10.0028193}{Perception and imitation
  of period-doubled phonation: {{Pitch}} and voice quality},'' \emph{J. Acoust.
  Soc. Am.}, vol. 156, no.~2, pp. 1391--1412, Aug. 2024.

\bibitem[Ikuma et~al.(2024)Ikuma, McWhorter, and Kunduk]{ikuma2024d}
\BIBentryALTinterwordspacing
T.~Ikuma, A.~J. McWhorter, and M.~Kunduk, ``Evaluation of machine-learning
  pitch estimation algorithms,'' in \emph{13th {{ICVPB}}}, Erlangen, Germany,
  Jul. 2024, pp. 28--9. [Online]. Available:
  \url{https://www.icvpb-2024.de/documents/8/Program_ICVPB2024.pdf}
\BIBentrySTDinterwordspacing

\bibitem[Ikuma et~al.(2025)Ikuma, Kunduk, and McWhorter]{ikuma2025}
T.~Ikuma, M.~Kunduk, and A.~J. McWhorter,
  ``\href{https://doi.org/10.48550/ARXIV.2501.04789}{Comparison of fundamental
  frequency estimators with subharmonic voice signals},'' 2025, arXiv preprint,
  https://arxiv.org/abs/2501.04789.

\bibitem[Titze(1994)]{titze1994}
\BIBentryALTinterwordspacing
I.~R. Titze, \emph{Workshop on {{Acoustic Voice Analysis}}: {{Summary
  Statement}}}.\hskip 1em plus 0.5em minus 0.4em\relax Denver, CO, USA:
  {National Center for Voice and Speech}, 1994. [Online]. Available:
  \url{ncvs.org/archive/freebooks/summary-statement.pdf}
\BIBentrySTDinterwordspacing

\bibitem[{KayPENTAX}(2008)]{kaypentax2008}
{KayPENTAX}, ``Multi-{{Dimensional Voice Program}} ({{MDVP}}) model 5105
  software instruction manual,'' Jun. 2008.

\bibitem[Deliyski(1993)]{deliyski1993}
D.~Deliyski, ``\href{https://doi.org/10.21437/Eurospeech.1993-445}{Acoustic
  model and evaluation of pathological voice production},'' in \emph{Eurospeech
  1993}, Berlin, Germany, 1993, pp. 969--1972.

\bibitem[Sun(2000)]{sun2000}
X.~Sun, ``\href{https://doi.org/10.21437/ICSLP.2000-902}{A pitch determination
  algorithm based on subharmonic-to-harmonic ratio},'' in \emph{Proc. 6th
  {{ICSLP}}}, vol.~4, Beijing, China, 2000, pp. 676--679.

\bibitem[Hlavni{\v c}ka et~al.(2019)Hlavni{\v c}ka, {\v C}mejla, Klemp{\'i}{\v
  r}, R{\r u}{\v z}i{\v c}ka, and Rusz]{hlavnicka2019}
J.~Hlavni{\v c}ka, R.~{\v C}mejla, J.~Klemp{\'i}{\v r}, E.~R{\r u}{\v z}i{\v
  c}ka, and J.~Rusz,
  ``\href{https://doi.org/10.1109/ACCESS.2019.2945874}{Acoustic tracking of
  pitch, modal, and subharmonic vibrations of vocal folds in {{Parkinson}}'s
  {{Disease}} and {{Parkinsonism}}},'' \emph{IEEE Access}, vol.~7, pp.
  150\,339--150\,354, 2019.

\bibitem[Aichinger et~al.(2017)Aichinger, Hagm{\"u}ller, Roesner,
  {Schneider-Stickler}, Schoentgen, and Pernkopf]{aichinger2017b}
P.~Aichinger, M.~Hagm{\"u}ller, I.~Roesner, B.~{Schneider-Stickler},
  J.~Schoentgen, and F.~Pernkopf,
  ``\href{https://doi.org/10.1016/j.bspc.2016.10.002}{Fundamental frequency
  tracking in diplophonic voices},'' \emph{Biomed. Signal Process. Control},
  vol.~37, pp. 69--81, Aug. 2017.

\bibitem[Aichinger et~al.(2018)Aichinger, Hagm{\"u}ller, {Schneider-Stickler},
  Schoentgen, and Pernkopf]{aichinger2018a}
P.~Aichinger, M.~Hagm{\"u}ller, B.~{Schneider-Stickler}, J.~Schoentgen, and
  F.~Pernkopf, ``\href{https://doi.org/10.1109/TASLP.2017.2761233}{Tracking of
  multiple fundamental frequencies in diplophonic voices},'' \emph{IEEEACM
  Trans. Audio Speech Lang. Process.}, vol.~26, no.~2, pp. 330--341, Feb. 2018.

\bibitem[Awan and Awan(2020)]{awan2020}
S.~N. Awan and J.~A. Awan,
  ``\href{https://doi.org/10.1016/j.jvoice.2018.07.003}{A two-stage cepstral
  analysis procedure for the classification of rough voices},'' \emph{J.
  Voice}, vol.~34, no.~1, pp. 9--19, Jan. 2020.

\bibitem[Kitayama et~al.(2023)Kitayama, Hosokawa, Iwaki, Yoshida, Miyauchi,
  Ogawa, and Inohara]{kitayama2023}
I.~Kitayama, K.~Hosokawa, S.~Iwaki, M.~Yoshida, A.~Miyauchi, M.~Ogawa, and
  H.~Inohara, ``\href{https://doi.org/10.1016/j.jvoice.2023.12.002}{Validation
  of subharmonics quantification using two-stage cepstral analysis},'' \emph{J.
  Voice}, p. S0892199723003892, Dec. 2023.

\bibitem[Ardaillon and Roebel(2019)]{ardaillon2019}
L.~Ardaillon and A.~Roebel,
  ``\href{https://doi.org/10.21437/Interspeech.2019-2815}{Fully-convolutional
  network for pitch estimation of speech signals},'' in \emph{Interspeech
  2019}, Sep. 2019, pp. 2005--2009.

\bibitem[Kim et~al.(2018)Kim, Salamon, Li, and Bello]{kim2018}
J.~W. Kim, J.~Salamon, P.~Li, and J.~P. Bello,
  ``\href{https://doi.org/10.1109/ICASSP.2018.8461329}{Crepe: A convolutional
  representation for pitch estimation},'' in \emph{{{IEEE ICASSP}} 2018},
  Calgary, AB, Apr. 2018, pp. 161--165.

\bibitem[Long et~al.(2015)Long, Shelhamer, and Darrell]{long2015}
J.~Long, E.~Shelhamer, and T.~Darrell,
  ``\href{https://doi.org/10.1109/CVPR.2015.7298965}{Fully convolutional
  networks for semantic segmentation},'' in \emph{Proc {{IEEE CVPR}}
  2015}.\hskip 1em plus 0.5em minus 0.4em\relax Boston, MA, USA: IEEE, Jun.
  2015, pp. 3431--3440.

\bibitem[Wang et~al.(2017)Wang, Yan, and Oates]{wang2017}
Z.~Wang, W.~Yan, and T.~Oates,
  ``\href{https://doi.org/10.1109/IJCNN.2017.7966039}{Time series
  classification from scratch with deep neural networks: {{A}} strong
  baseline},'' in \emph{Proc. {{IJCNN}} 2017}.\hskip 1em plus 0.5em minus
  0.4em\relax Anchorage, AK, USA: IEEE, May 2017, pp. 1578--1585.

\bibitem[Ismail~Fawaz et~al.(2019)Ismail~Fawaz, Forestier, Weber, Idoumghar,
  and Muller]{ismailfawaz2019}
H.~Ismail~Fawaz, G.~Forestier, J.~Weber, L.~Idoumghar, and P.-A. Muller,
  ``\href{https://doi.org/10.1007/s10618-019-00619-1}{Deep learning for time
  series classification: A review},'' \emph{Data Min. Knowl. Discov.}, vol.~33,
  no.~4, pp. 917--963, Jul. 2019.

\bibitem[Dejonckere and Lebacq(1983)]{dejonckere1983}
P.~Dejonckere and J.~Lebacq,
  ``\href{https://doi.org/10.1016/0167-6393(83)90063-8}{An analysis of the
  diplophonia phenomenon},'' \emph{Speech Commun.}, vol.~2, no.~1, pp. 47--56,
  May 1983.

\bibitem[Herbst(2021)]{herbst2021}
C.~T. Herbst, ``\href{https://doi.org/10.1016/j.jvoice.2019.11.005}{Performance
  evaluation of subharmonic-to-harmonic ratio ({{SHR}}) computation},''
  \emph{J. Voice}, vol.~35, no.~3, pp. 365--375, May 2021.

\bibitem[Titze(1984)]{titze1984}
I.~R. Titze, ``\href{https://doi.org/10.1121/1.390530}{Parameterization of the
  glottal area, glottal flow, and vocal fold contact area},'' \emph{J. Acoust.
  Soc. Am.}, vol.~75, no.~2, pp. 570--580, 1984.

\bibitem[Titze(1989)]{titze1989}
------, ``\href{https://doi.org/10.1016/0167-6393(89)90001-0}{A four-parameter
  model of the glottis and vocal fold contact area},'' \emph{Speech Commun.},
  vol.~8, no.~3, pp. 191--201, Sep. 1989.

\bibitem[Story(1995)]{story1995a}
B.~Story, ``Physiologically-{{Based Speech Simulation Using}} an {{Enhanced
  Wave-Reflection Model}} of the {{Vocal Tract}},'' Ph.D. dissertation,
  University of Iowa, Iowa City, IA, May 1995.

\bibitem[Liljencrants(1985)]{liljencrants1985}
J.~Liljencrants, ``Speech {{Synthesis}} with {{Reflection-Type Line Analog}},''
  Ph.D. dissertation, Royal Institute of Technology, Stockholm, Sweden, 1985.

\bibitem[Samlan and Story(2011)]{samlan2011}
R.~A. Samlan and B.~H. Story,
  ``\href{https://doi.org/10.1044/1092-4388(2011/10-0195)}{Relation of
  structural and vibratory kinematics of the vocal folds to two acoustic
  measures of breathy voice based on computational modeling},'' \emph{J. Speech
  Lang. Hear. Res.}, vol.~54, no.~5, pp. 1267--1283, Oct. 2011.

\bibitem[Samlan et~al.(2013)Samlan, Story, and Bunton]{samlan2013}
R.~A. Samlan, B.~H. Story, and K.~Bunton,
  ``\href{https://doi.org/10.1044/1092-4388(2012/12-0194)}{Relation of
  perceived breathiness to laryngeal kinematics and acoustic measures based on
  computational modeling},'' \emph{J. Speech Lang. Hear. Res.}, vol.~56, no.~4,
  pp. 1209--1223, Aug. 2013.

\bibitem[Samlan et~al.(2014)Samlan, Story, Lotto, and Bunton]{samlan2014a}
R.~A. Samlan, B.~H. Story, A.~J. Lotto, and K.~Bunton,
  ``\href{https://doi.org/10.1044/2014_JSLHR-S-12-0405}{Acoustic and perceptual
  effects of left--right laryngeal asymmetries based on computational
  modeling},'' \emph{J. Speech Lang. Hear. Res.}, vol.~57, no.~5, pp.
  1619--1637, Oct. 2014.

\bibitem[Ishizaka and Flanagan(1972)]{ishizaka1972}
K.~Ishizaka and J.~L. Flanagan,
  ``\href{https://doi.org/10.1002/j.1538-7305.1972.tb02651.x}{Synthesis of
  voiced sounds from a two-mass model of the vocal cords},'' \emph{Bell Syst.
  Tech. J.}, vol.~51, no.~6, pp. 1233--1268, 1972.

\bibitem[Titze(1988)]{titze1988}
I.~R. Titze, ``\href{https://doi.org/10.1121/1.395910}{The physics of
  small-amplitude oscillation of the vocal folds},'' \emph{J. Acoust. Soc.
  Am.}, vol.~83, no.~4, pp. 1536--1552, 1988.

\bibitem[Kiritani et~al.(1993)Kiritani, Hirose, and Imagawa]{kiritani1993}
S.~Kiritani, H.~Hirose, and H.~Imagawa,
  ``\href{https://doi.org/10.1016/0167-6393(93)90056-q}{High-speed digital
  image analysis of vocal cord vibration in diplophonia},'' \emph{Speech
  Commun.}, vol.~13, no. 1-2, pp. 23--32, 1993.

\bibitem[Kniesburges et~al.(2016)Kniesburges, Lodermeyer, Becker, Traxdorf, and
  D{\"o}llinger]{kniesburges2016}
S.~Kniesburges, A.~Lodermeyer, S.~Becker, M.~Traxdorf, and M.~D{\"o}llinger,
  ``\href{https://doi.org/10.1121/1.4954264}{The mechanisms of subharmonic tone
  generation in a synthetic larynx model},'' \emph{J. Acoust. Soc. Am.}, vol.
  139, no.~6, pp. 3182--3192, Jun. 2016.

\bibitem[Ikuma et~al.(2016)Ikuma, Kunduk, Fink, and McWhorter]{ikuma2016a}
T.~Ikuma, M.~Kunduk, D.~Fink, and A.~J. McWhorter,
  ``\href{https://doi.org/10.1121/1.4964400}{Synthetic multi-line kymographic
  analysis: {{A}} spatiotemporal data reduction technique for high-speed
  videoendoscopy},'' \emph{J. Acoust. Soc. Am.}, vol. 140, no.~4, pp.
  2703--2713, Oct. 2016.

\bibitem[Mergell et~al.(2000)Mergell, Herzel, and Titze]{mergell2000}
P.~Mergell, H.~Herzel, and I.~R. Titze,
  ``\href{https://doi.org/10.1121/1.1314398}{Irregular vocal-fold
  vibration---{{High-speed}} observation and modeling},'' \emph{J. Acoust. Soc.
  Am.}, vol. 108, no.~6, pp. 2996--3002, 2000.

\bibitem[Neubauer et~al.(2001)Neubauer, Mergell, Eysholdt, and
  Herzel]{neubauer2001}
J.~Neubauer, P.~Mergell, U.~Eysholdt, and H.~Herzel,
  ``\href{https://doi.org/10.1121/1.1406498}{Spatio-temporal analysis of
  irregular vocal fold oscillations: {{Biphonation}} due to desynchronization
  of spatial modes},'' \emph{J. Acoust. Soc. Am.}, vol. 110, no.~6, pp.
  3179--3192, 2001.

\bibitem[Titze(2002)]{titze2002}
I.~R. Titze, ``\href{https://doi.org/10.1121/1.1417526}{Regulating glottal
  airflow in phonation: {{Application}} of the maximum power transfer theorem
  to a low dimensional phonation model},'' \emph{J. Acoust. Soc. Am.}, vol.
  111, no.~1, pp. 367--376, Jan. 2002.

\bibitem[Klatt and Klatt(1990)]{klatt1990}
D.~H. Klatt and L.~C. Klatt,
  ``\href{https://doi.org/10.1121/1.398894}{Analysis, synthesis, and perception
  of voice quality variations among female and male talkers},'' \emph{J.
  Acoust. Soc. Am.}, vol.~87, no.~2, pp. 820--857, 1990.

\bibitem[Holmberg et~al.(1988)Holmberg, Hillman, and Perkell]{holmberg1988}
E.~B. Holmberg, R.~E. Hillman, and J.~S. Perkell,
  ``\href{https://doi.org/10.1121/1.396829}{Glottal airflow and transglottal
  air pressure measurements for male and female speakers in soft, normal, and
  loud voice},'' \emph{J. Acoust. Soc. Am.}, vol.~84, no.~2, pp. 511--529, Aug.
  1988.

\bibitem[Kingma and Ba(2015)]{kingma2015}
D.~P. Kingma and J.~Ba, ``\href{https://doi.org/10.48550/arXiv.1412.6980}{Adam:
  {{A Method}} for {{Stochastic Optimization}}},'' in \emph{{{ICLR}}
  2015}.\hskip 1em plus 0.5em minus 0.4em\relax San Diego, CA: arXiv, 2015.

\bibitem[{KayPENTAX} and {Massachusetts Eye and Ear
  Infirmary}(2006)]{kaypentax2006}
{KayPENTAX} and {Massachusetts Eye and Ear Infirmary}, ``Disordered {{Voice
  Database}} and {{Program}} [{{Model}} 4337],'' 2006.

\end{thebibliography}
\end{document}